\documentclass{aa}
\usepackage{epsfig}

\newbox\grsign \setbox\grsign=\hbox{$>$}
\newdimen\grdimen \grdimen=\ht\grsign
\newbox\laxbox \newbox\gaxbox
\setbox\gaxbox=\hbox{\raise.5ex\hbox{$>$}\llap
     {\lower.5ex\hbox{$\sim$}}}\ht1=\grdimen\dp1=0pt
\setbox\laxbox=\hbox{\raise.5ex\hbox{$<$}\llap
     {\lower.5ex\hbox{$\sim$}}}\ht2=\grdimen\dp2=0pt

\def\lesssim{\mathrel{\hbox{\rlap{\hbox{\lower4pt\hbox{$\sim$}}}\hbox{$<$}}}}
\def\gtrsim{\mathrel{\hbox{\rlap{\hbox{\lower4pt\hbox{$\sim$}}}\hbox{$>$}}}}

\begin{document}



\title{SCUBA sub-millimeter observations of gamma-ray bursters}

\subtitle{II. GRB 991208, 991216, 000301C, 000630, 000911, 000926}

\author{I.A. Smith\inst{1}
\and R.P.J. Tilanus\inst{2}
\and R.A.M.J. Wijers\inst{3}
\and N. Tanvir\inst{4}
\and P. Vreeswijk\inst{5}
\and E. Rol\inst{5}
\and C. Kouveliotou\inst{6,7}}

\institute{Department of Physics and Astronomy,
Rice University, MS-108, 6100 South Main, Houston, TX 77005-1892 USA
\and Joint Astronomy Centre, 660 N. Aohoku Place, Hilo, HI 96720 USA
\and Department of Physics and Astronomy, SUNY,
Stony Brook, NY 11794-3800 USA
\and Department of Physical Sciences, University of Hertfordshire,
College Lane, Hatfield, Herts AL10 9AB, UK
\and Astronomical Institute `Anton Pannekoek', 
University of Amsterdam and Center for High-Energy Astrophysics,\\
Kruislaan 403, 1098 SJ Amsterdam, The Netherlands
\and NASA Marshall Space Flight Center, SD-50, NSSTC, 
320 Sparkman Drive, Huntsville, AL 35805 USA
\and Universities Space Research Association}

\offprints{Ian A. Smith}

\date{Received 31 January 2001 / Accepted 2 October 2001}

\abstract{We discuss our ongoing program of Target of Opportunity 
sub-millimeter observations of gamma-ray bursts (GRBs) 
using the Sub-millimetre Common-User Bolometer Array (SCUBA) 
on the James Clerk Maxwell Telescope (JCMT).
Sub-millimeter observations of the early afterglows are of interest
because this is where the emission peaks in some bursts in the days 
to weeks following the burst.
Of increasing interest is to look for underlying quiescent sub-millimeter
sources that may be dusty star-forming host galaxies.
In this paper, we present observations of GRB 991208, 991216, 000301C, 
000630, 000911, and 000926.
For all these bursts, any sub-millimeter emission is consistent with coming 
from the afterglow.
This means that we did not conclusively detect quiescent sub-millimeter
counterparts to any of the bursts that were studied from 1997 through
2000.
The inferred star formation rates ($M \ge 5 M_{\sun}$) are typically 
$\lesssim 300 ~M_{\sun} {\rm yr}^{-1}$.
If GRBs are due to the explosions of high-mass stars, this may indicate 
that the relatively small population of extremely luminous dusty galaxies 
does not dominate the total star formation in the universe at early epochs.
Instead, the GRBs may be predominantly tracing slightly lower luminosity 
galaxies.
The optical faintness of some host galaxies is unlikely to be explained 
as due to dust absorption in the host.
\keywords{gamma rays: bursts -- radio continuum: general -- infrared: general}
}

\maketitle

\section{Introduction}

The discovery of localized transients in the error boxes of
gamma-ray burst (GRB) sources has led to intense multi-wavelength
campaigns that have revolutionized our understanding of these sources.
For reviews, see Piran (\cite{scuba:piran99});
Van Paradijs et al. (\cite{scuba:vanp00}).

The multi-wavelength emission comes from several components.
The ``prompt'' emission comes from the initial explosion.
A reverse shock can give an optical and/or radio flash.
The later ``afterglow'' emission comes from the expanding fireball as it 
sweeps up the surrounding medium.
Finally, the ``quiescent'' constant emission comes from any underlying 
host galaxy.

Sub-millimeter observations provide ``clean'' measures of the source
intensity, unaffected by scintillation and extinction.
We have been performing Target of Opportunity 
sub-millimeter observations of GRB counterparts using the 
Sub-millimetre Common-User Bolometer Array (SCUBA) 
on the James Clerk Maxwell Telescope (JCMT) on Mauna Kea, Hawaii.
The detailed SCUBA results for the first eight (GRB 970508, 971214, 
980326, 980329, 980519, 980703, 981220, and 981226) are described in
Smith et al. (\cite{scuba:smith99}).
GRB 990123 is discussed in Galama et al. (\cite{scuba:gal99nat}).
Observations of GRB 990520 were made in mediocre weather
(Smith et al. \cite{scuba:smith00}).
In this paper we present our SCUBA observations of 
GRB 991208, 991216, 000301C, 000630, 000911, and 000926.
The sub-millimeter emission from GRB 010222 that was detected 
by SCUBA appeared to have unusual properties 
(Fich et al. \cite{scuba:fich01}; Ivison et al. \cite{scuba:ivi01});
these observations will be reported elsewhere after a final sensitive SCUBA
observation has been made to determine whether there is an 
underlying host galaxy.

In the remainder of \S 1, we outline the motivations for making
sub-millimeter observations of the afterglows and host galaxies.
In \S 2 we discuss the technical details of the SCUBA observations,
and also the chance probability of detecting quiescent sources that are 
unrelated to the GRB.
In \S 3 we present the results of our latest observations.
In \S 4 we discuss some of the implications of our results.

\subsection{SCUBA afterglow program}

Both observations (e.g. Galama et al. \cite{scuba:galama98},
\cite{scuba:gal99nat}; Smith et al. \cite{scuba:smith99}) and theories 
(e.g. Sari et al. \cite{scuba:spn98}; Piran \cite{scuba:piran99};
Wijers \& Galama \cite{scuba:wg99}; Granot et al. \cite{scuba:gps00};
Sari \& M\'esz\'aros \cite{scuba:sm00}; Chevalier \& Li \cite{scuba:cl00};
Panaitescu \& Kumar \cite{scuba:pk00})
show that the burst afterglow emission often peaks in the sub-millimeter in 
the days to weeks following the burst.
By tracking the evolving emission across the entire spectrum, 
it is possible to study aspects such as the types of shocks involved,
the geometry of the outflow (jet versus spherical), and the geometry of
the surrounding medium (uniform versus stellar wind).

Previously, our most interesting result was the detection of a fading 
sub-millimeter counterpart to GRB 980329 (Smith et al. \cite{scuba:smith99}).
While a fading X-ray counterpart (in't Zand et al. \cite{scuba:intz98})
and a variable, long-lasting radio source VLA J070238.0+385044 (Taylor et al. 
\cite{scuba:tay98}) were soon found for this burst, it proved to be 
difficult to find the infrared counterpart (Klose et al. \cite{scuba:klose98};
Palazzi et al. \cite{scuba:pal98}; Reichart et al. \cite{scuba:reich99}).
However, the 850 $\mu$m (350 GHz) flux seen by SCUBA was relatively bright, 
making it similar to GRB 970508.
The radio through sub-millimeter spectrum of GRB 980329 was well fit by a
power law with a surprisingly steep index of $\alpha = +0.9$ (Taylor et al. 
\cite{scuba:tay98}; Smith et al. \cite{scuba:smith99}).
However, we could not fully exclude a $\nu^{1/3}$ power law attenuated by 
synchrotron self-absorption (e.g. Katz \cite{scuba:katz94}; 
Waxman \cite{scuba:wax97}).
The 850 $\mu$m SCUBA flux decayed rapidly with time.
For a power law decay with the flux density $\propto t^{-m}$ where $t$ was 
the time since the burst, the best fit power law index was $m = 3.0$.
However, $m$ was not tightly constrained: the 90\% confidence interval was
$m = 1.2$ to $m = 5.3$.

\subsection{SCUBA host galaxy program}

It is important to understand the nature of the host galaxies.
A popular model for GRBs is that they are due to the explosive deaths of
high-mass stars (e.g. Woosley \cite{scuba:woosley93}, \cite{scuba:woosley00};
Paczy\'nski \cite{scuba:pac98}) and therefore are likely to be found in 
active star forming regions.
SCUBA has recently discovered several dusty star-forming galaxies out to high
redshifts (e.g. Smail et al. \cite{scuba:smail97}, \cite{scuba:smail98}; 
Hughes et al. \cite{scuba:hughes98}; Barger et al. \cite{scuba:barger98}), 
and it appears that the star formation rate does not drop rapidly beyond 
$z \sim 1$ (e.g. Blain et al. \cite{scuba:blain99}).
In this scenario, it is plausible that some of the GRBs will
be associated with dusty star-forming galaxies.

An alternative model assumes that GRBs are due to the coalescence of two 
compact objects that were in a binary orbit.
In this case, the lifetime of the binary will likely be long, and the 
active star formation will have died down by the time the burst occurs.
Furthermore, the binary system may have obtained a substantial velocity
if there was a supernova explosion with a large kick; this
would move the system far from its birthplace.
Thus GRBs would be less likely to occur in a dusty star-forming galaxy.

If there is a connection between GRBs and dust-enshrouded star formation
(Blain \& Natarajan \cite{scuba:bn00}), it has been suggested that 
$\gtrsim 20$\% of GRB hosts should be brighter than 2 mJy at 
850 $\mu$m (Ramirez-Ruiz et al. \cite{scuba:rtb01}).
Our Target of Opportunity program is designed to look for the afterglow 
emission of the bursts by making observations separated by days.
However, by combining the data from all of our observations of a source 
we can also look for quiescent sources of this nature.

The expected sub-millimeter emission from warm dust due to massive star 
formation is somewhat uncertain.
In this paper, we will give numerical estimates using 
(Condon \cite{scuba:condon}; Carilli \& Yun \cite{scuba:cy99}):
\begin{eqnarray}
S_\nu = 0.039~ \xi(z, q_0, \alpha) &
{\left[ H_0 \over 65~{\rm kms}^{-1}{\rm Mpc}^{-1} \right]}^2 
{\left[ \nu \over 350~{\rm GHz} \right]}^\alpha \nonumber \\
 & \times
{\left[ {{\rm SFR}~(M \ge 5 M_{\sun} )} \over {100 ~M_{\sun}
{\rm yr}^{-1}} \right]} ~{\rm mJy}\,,
\end{eqnarray}
where $S_\nu$ is the sub-millimeter flux density at the frequency $\nu$.

SFR is the star formation rate for stars with masses $\ge 5 M_{\sun}$ in
units of $100 ~M_{\sun} {\rm yr}^{-1}$.
An advantage of using the sub-millimeter flux density to estimate
the SFR is that there is no need to perform the large (sometimes orders 
of magnitude) correction for the host galaxy extinction that is required 
when studying optical lines.
Note that other methods of estimating the SFR often quote results
using masses between 0.1 and 100 $M_{\sun}$.
Our results will be smaller than these by a factor that depends on the
initial mass function (IMF).
This is currently uncertain for starburst galaxies; 
for a Salpeter IMF, the factor is $\sim 5$.

The function $\xi(z, q_0, \alpha)$ is given by:
\begin{equation}
\xi(z, q_0, \alpha) = {(1+z)^{1+\alpha} \over 
z^2 \left[ 1 + {z(1-q_0) \over {(1+2 q_0 z)^{1/2} + 1 + q_0 z}} \right]^2}
\,.
\end{equation}
This assumes that the Universe has a geometry described by the
Robertson-Walker metric and uses the Friedmann cosmological models.
Typically, $\xi(z, q_0, \alpha) \sim 10$ over a wide range of 
redshifts, assuming plausible values for $q_0$ and $\alpha$.
In our numerical calculations, we assume
$H_0 = 65~{\rm kms}^{-1}{\rm Mpc}^{-1}$,
$\Omega_0 = 0.2$, $q_0 = 0.1$, and $\Lambda_0 = 0$.

The power law index $\alpha$ is measured between 230 and 850 GHz, and
is expected to be $\sim 3 - 4$ based on observations and simple dust models 
(e.g. Dwek \& Werner \cite{scuba:dwek81}).
Thus any quiescent dust contribution is very much larger at sub-millimeter 
than at radio wavelengths.
In principle, $\alpha$ could be measured by SCUBA.
However, in practice the weather is rarely good enough to be able to
detect the source at 450 $\mu$m.
For our GRB numerical calculations, we assume $\alpha = 3.4$
(which is the value observed for M82).
This lies in the middle of the expected range of $\alpha$, and for the
$z \lesssim 2$ bursts considered here the uncertainty in $\alpha$ means
that the calculated SFR may be too high or too low by at most 50\%.
Note that depending on the temperature of the dust, for very large 
$z \gtrsim 7$ the peak of the dust emitting spectrum may be redshifted 
to a wavelength $\sim 850~ \mu$m; in this case, $\alpha$ will be much 
flatter.

As an example, a starburst galaxy with a 
${\rm SFR} = 300 ~M_{\sun} {\rm yr}^{-1}$, and $\alpha = 3.0$
(as in Arp 220) would give an 850 $\mu$m flux density of
$S_\nu = 1.5~{\rm mJy}$ at $z = 10$.
Alternatively, a flux density of $S_\nu = 1.5~{\rm mJy}$ at 850 $\mu$m
for an ultraluminous infrared galaxy (ULIRG) at $z=1$ with $\alpha = 3.4$ 
would imply a ${\rm SFR} = 360 ~M_{\sun} {\rm yr}^{-1}$.

SCUBA is therefore capable of detecting ULIRG-type objects throughout 
the distant universe.
The relatively small population of extremely luminous dusty
galaxies may dominate the total star formation in the universe at
early epochs (Smail et al. \cite{scuba:smail00};
Ramirez-Ruiz et al. \cite{scuba:rtb01}). 
However, this is still quite uncertain; for example, there may be a
substantial contribution to the energy output of the SCUBA-bright
galaxies from active galactic nuclei.
If GRBs are regularly found to be associated with extremely luminous 
dusty galaxies, this will provide independent evidence that these 
objects dominate the star formation in the early universe.

In Smith et al. (\cite{scuba:smith99}) we noted that the relatively
bright sub-millimeter flux from GRB 980329 could be due in part to an
underlying quiescent sub-millimeter source.
Unfortunately, the quiescent flux of $\sim 1$ mJy at 850 $\mu$m in this 
case is too faint to be significantly detected by SCUBA: more sensitive 
sub-millimeter instruments will be needed to determine whether such a 
quiescent source is actually present in GRB 980329.

None of the other GRBs presented in our previous papers showed any 
indication of an underlying quiescent sub-millimeter source.
We discuss the SFR limits for these in \S 4.
However, 60 $\mu$m {\it Infrared Space Observatory} observations 
suggested a possible far-infrared quiescent counterpart to GRB 970508 
(Hanlon et al. \cite{scuba:hanlon00}).

\section{SCUBA details}

SCUBA is the sub-millimeter continuum instrument for the JCMT
(for a review see Holland et al. \cite{scuba:hol99}).
Our observing, calibration, and reduction techniques are the same as 
described in detail in Smith et al. (\cite{scuba:smith99}).
Here we summarize only the most important features of the instrument, and 
those aspects that have been modified since our previous observations.
We also discuss the chance probability of detecting a quiescent 
source that is unrelated to the GRB.

\subsection{SCUBA observing details}

SCUBA uses two arrays of bolometers to simultaneously observe the same
region of sky, $\sim 2.3\arcmin$ in diameter.
The arrays are optimized for operations at 450 and 850 $\mu$m.
The dedicated photometry pixels for observations at 1100, 1350, and 
2000 $\mu$m have not been available since the 1999 upgrade.

In 1999 October new wideband filters (450w:850w) were installed.
These have improved sensitivities over the ones used previously by SCUBA.
New blocking filters with better transmission were also installed in 
1999 October.
New ribbon cables now allow the refrigerator to run colder: 
this has reduced the 
number of noisy pixels and has stabilized the noise on the arrays.
This updated set-up was used for all the new observations described here,
and has generally resulted in lower rms values than was found in 
Smith et al. (\cite{scuba:smith99}).
A typical integration time of 2 hours gives an rms $\sim 1.5$ mJy at 
850 $\mu$m.
However, the sensitivity depends significantly on the weather and
the elevation of the source; since our ToO observations are done
on short notice, sometimes these factors are less than ideal.

Fully sampled maps of the $2.3\arcmin$ region can be made by ``jiggling'' 
the array.
However, for all the sources described here, we have only been looking at the
well-localized optical or radio transient coordinates by performing deep
photometry observations using a single pixel of the arrays.
The other bolometers in the arrays are used to perform a good sky noise 
subtraction.

During an observation the secondary is chopped between the
source and sky at 7 Hz.
This is done mainly to take out small relative DC drifts between the
bolometers, and also to remove any large-scale sky variations.
The term ``integration time'' always refers to the ``on+off''
time, including the amount of time spent while chopped off-source.
An 18 sec integration thus amounts to a 9 sec on-source observation time.
A typical measurement consists of 50 integrations of 18 seconds;
we refer to this as a ``run.''
Each observation of a source in general consists of several such
runs, with pointing and calibration observations in between.

Based on observed variations of the gain factor and signal levels we estimate 
typical systematic uncertainties in the absolute flux calibrations of 10\% at 
850 $\mu$m.
In general the rms errors of the observations presented here are larger than 
this uncertainty.

\subsection{False positives}

The pointing accuracy of the JCMT is a few arcsec, and the pointing is 
checked several times during the night to ensure that it is reliable.
The 850 $\mu$m bolometric pixel has a diffraction limited resolution
of $14\arcsec$.
Thus we can be sure that the target is always well centered in the 
bolometric pixel.

However, the large beam size combined with the large number of 
distant galaxies radiating strongly at this wavelength means that
in any observation there is a non-negligible chance of detecting a 
quiescent sub-millimeter source that is completely unrelated to the GRB.

The surface density of sub-millimeter galaxies is still somewhat uncertain
(Blain et al. \cite{scuba:blain98};
Barger et al. \cite{scuba:barger99}).
At 850 $\mu$m, the surface density of galaxies with flux densities larger 
than 4 mJy is estimated to be $\sim (1 - 2.5) \times 10^3~{\rm deg}^{-2}$, 
while the surface density of galaxies with flux densities larger than 
1 mJy is $\gtrsim 10^4~{\rm deg}^{-2}$.

Because of the uncertainty in the surface density, and the fact that
the sensitivity of the bolometric pixel depends on the off-axis
distance of the source, it is difficult to give an accurate determination 
of the chance probability of detecting a quiescent sub-millimeter source 
that is completely unrelated to the GRB.
However, we estimate that the chance of detecting a random $\ge 4$ mJy 
source in any pointing is $\sim 1 - 3$\%, while the chance of detecting a 
random $\ge 1$ mJy source in any pointing is $> 10$\%.

The rather large chance of a false positive means that it is dangerous
to use just the detection of a quiescent sub-millimeter source to claim
that this must be the host galaxy to the GRB.
For example, we cannot rule out that the 1 mJy source
that may have been seen in GRB 980329 is an unrelated source in
the JCMT beam.
For individual cases, confirmation of the star formation rate is
needed from observations at other wavelengths.
However, if it is found that many more bursts are associated with 
quiescent sub-millimeter sources than is expected by chance, this would 
be good evidence that the majority of these are true associations.

\section{Results of SCUBA observations}

The fact that SCUBA was down during the 1999 upgrade combined with
unusually poor weather and a lack of well-localized sources in regions 
of the sky accessible to SCUBA has unfortunately limited our program over 
the past couple of years.
For completeness, we describe here all of our observations from late
1999 through 2000.
Table 1 summarizes the 850 $\mu$m results.

\begin{table}[t]
\caption[]{SCUBA 850 $\mu$m (353 GHz) GRB observations.}
\label{table1}
\begin{flushleft}
\[
\begin{tabular}{lllll}
\hline
\noalign{\smallskip}
Burst   & $z$   & Observing   & Time since       & 850 $\mu$m flux  \\
        &       & date        & burst            & density          \\
        &       & (UT)        & (days)           & (mJy)            \\
\noalign{\smallskip}
\hline
\noalign{\smallskip}
991208  & 0.707 & 19991215.82 & \phantom{1}7.63  & $\phantom{-} 3.4 \pm 3.7$ \\
        &       & 19991219.82 &           11.63  &            $-0.8 \pm 1.8$ \\
\noalign{\medskip}
991216  & 1.0   & 19991218.48 & \phantom{1}1.81  & $\phantom{-} 0.7 \pm 1.6$ \\
        &       & 19991219.45 & \phantom{1}2.78  &            $-2.0 \pm 1.7$ \\
\noalign{\medskip}
000301C & 2.04  & 20000304.75 & \phantom{1}3.34  & $\phantom{-} 3.1 \pm 3.1$ \\
        &       & 20000305.53 & \phantom{1}4.12  & 
$\phantom{-} 1.9 \pm 1.2^{\mathrm{a}}$ \\
        &       & 20000306.50 & \phantom{1}5.09  & $\phantom{-} 1.1 \pm 0.9$ \\ 
\noalign{\medskip}
000911  &       & 20000917.53 & \phantom{1}6.23  &            $-0.4 \pm 1.4$ \\
        &       & 20000920.49 & \phantom{1}9.19  & $\phantom{-} 0.3 \pm 1.1$ \\
\noalign{\medskip}
000926  & 2.04  & 20000930.27 & \phantom{1}3.28  & $\phantom{-} 7.3 \pm 4.2$ \\
\noalign{\smallskip}
\hline
\end{tabular}
\]
\end{flushleft}
\begin{list}{}{}
\item[$^{\mathrm{a}}$]This result uses all of our data, including a 
negative outlier.
Alternatively, we could quote a ``best'' result of $3.3 \pm 1.2$ mJy 
around the time of the peak magnification in the lensing scenario.
See the text for details.
\end{list}
\end{table}

\subsection{GRB 991208}

The extremely energetic GRB 991208 was detected 19991208.19 
(Hurley et al. \cite{scuba:hur00a}) 
by the {\it Ulysses}, {\it Near-Earth Asteroid Rendezvous} ({\it NEAR}), and
{\it Wind} spacecraft of the third Interplanetary Network (IPN).

A radio source was found by the Very Large Array (VLA: 
Frail \cite{scuba:frail99}; Hurley et al. \cite{scuba:hur00a}).
This source had a rising spectral index.
A few days after the burst, the source was found to have fluxes of a couple
of mJy from 15 to 240 GHz using the Ryle Telescope 
(Pooley \cite{scuba:pooley99a}), Owens Valley Millimeter Array
(Shepherd et al. \cite{scuba:shep99}), and IRAM (Bremer et al.
\cite{scuba:bremer99}).

An optical transient was also found (Castro-Tirado et al. \cite{scuba:ct99};
Stecklum et al. \cite{scuba:stecklum99}) whose optical flux decayed
following a power law $\propto t^{-\delta}$ with a steep slope 
$\delta = 2.3 \pm 0.07$ for the first $\sim 5$ days, breaking to an
even faster $\delta = 3.2 \pm 0.2$ (Castro-Tirado et al. \cite{scuba:ct01}).

A redshift of $z = 0.707$ was determined for the afterglow 
(Dodonov et al. \cite{scuba:dod99}; Djorgovski et al. \cite{scuba:djor99};
Castro-Tirado et al. \cite{scuba:ct01}; Sokolov et al. \cite{scuba:sok01}).
There was an indication of a supernova or dust echo emission component
in the afterglow (Castro-Tirado et al. \cite{scuba:ct01}).
A compact $V = 24.6$ magnitude galaxy was later found at this location 
(Fruchter et al. \cite{scuba:fruchter00}).

Unfortunately, GRB 991208 was in the morning sky and during a period of
very poor weather at Mauna Kea.
Our first opportunity to make a (short) observation was not until a week 
after the burst.
Table 1 shows the results of our two observations.
They are consistent with the other millimeter results
(Galama et al. \cite{scuba:gal00}), though we are 
unable to determine anything significant about the evolution of the flux.

A SFR of $(11.5 \pm 7.1) ~M_{\sun} {\rm yr}^{-1}$ was estimated
from the emission line fluxes of the host galaxy 
(Castro-Tirado et al. \cite{scuba:ct01}).
The expected quiescent flux density at 850 $\mu$m would then be
$S_\nu \sim 0.06~{\rm mJy}$, assuming $\alpha = 3.4$.
However, correcting for the internal extinction in the host 
galaxy leads to significantly higher SFR estimates,
$>156 ~M_{\sun} {\rm yr}^{-1}$ to $>249 ~M_{\sun} {\rm yr}^{-1}$
(Sokolov et al. \cite{scuba:sok01}).
A SFR ($M \ge 5 M_{\sun}$) of $250 ~M_{\sun} {\rm yr}^{-1}$ would give 
a flux density of $\sim 1.2$ mJy at 850 $\mu$m;
since we do not detect a quiescent sub-millimeter source with
SCUBA, the SFR cannot be significantly higher than this value.

\subsection{GRB 991216}

The extremely bright GRB 991216 was detected 19991216.67 
by BATSE on the {\it Compton Gamma-Ray Observatory} ({\it CGRO})
(Kippen et al. \cite{scuba:kpg99}; Kippen \cite{scuba:kip99}), 
and by {\it Ulysses}, {\it NEAR}, and the 
{\it Satellite per Astronomia a Raggi X} ({\it BeppoSAX})
(Hurley \& Feroci \cite{scuba:hf99};
Hurley et al. \cite{scuba:hck99}).
Scans of the GRB error box using the Proportional Counter Array 
({\it PCA}) on the {\it Rossi X-ray Timing Explorer} ({\it RXTE})
discovered a previously unknown fading X-ray source
(Takeshima et al. \cite{scuba:tak99}).
This source was accurately located by the {\it Chandra X-Ray Observatory} 
({\it CXO}) (Piro et al. \cite{scuba:piro99}).

The X-ray transient location was consistent with that of a fading optical 
source (Uglesich et al. \cite{scuba:ug99}), whose optical power law decay 
of $\delta \sim 1.2 - 1.5$ was flatter than the X-ray decay of $\sim 1.6$
(Garnavich et al. \cite{scuba:garn00}; Halpern et al. \cite{scuba:halp00}).
Detections of both redshifted iron features
(Piro et al. \cite{scuba:pgg00}) and optical absorption lines
(Vreeswijk et al. \cite{scuba:vrees99})
indicated that the burst was at $z = 1.0$.
An irregularly shaped $R = 26.9$ magnitude galaxy was later found at this 
location (Vreeswijk et al. \cite{scuba:vrees00}).

A variable radio afterglow was also found soon after the burst
(Taylor \& Berger \cite{scuba:tb99}; Rol et al. \cite{scuba:rol99}).
Surprisingly, the radio decay started within 1.5 days of the burst, and
the radio power law decay index was less steep than in both the optical 
and X-ray bands (Frail et al. \cite{scuba:frail00}).
While a $1.1 \pm 0.25$ mJy source at 15 GHz was seen 1.33 days after the 
burst by the Ryle Telescope (Pooley \cite{scuba:pooley99b}), this was not 
detected in later Ryle observations, nor was it detected at 99.9 GHz by 
Owens Valley (Frail et al. \cite{scuba:frail00}).

Unfortunately, as with GRB 991208, the weather at Mauna Kea was very poor
during this period.
Our SCUBA results are listed in Table 1.
Our upper limits are consistent with the lack of a detection by Owens Valley.
We find no evidence for a quiescent dusty host galaxy for GRB 991216.
Combining all our 850 $\mu$m runs, the rms value of 1.2 mJy implies a 
SFR ($M \ge 5 M_{\sun}$) $< 290 ~M_{\sun} {\rm yr}^{-1}$, assuming 
$\alpha = 3.4$.

\subsection{GRB 000301C}

The short or intermediate duration ($2~{\rm s}$ in the $> 25~{\rm keV}$ 
energy range) GRB 000301C was detected 20000301.41 by the All Sky Monitor 
(ASM) on {\it RXTE}, by {\it Ulysses}, and by {\it NEAR} 
(Smith et al. \cite{scuba:dsmith00}; 
Jensen et al. \cite{scuba:jfg01}).

An optical counterpart was soon found in the burst error box
(e.g. Fynbo et al. \cite{scuba:fynbo00a};
Jensen et al. \cite{scuba:jfg01}).
The light curve of this relatively bright optical source was well 
sampled, and it was found that it had a complex behavior, with 
significant fluctuations superimposed on the overall decay
(e.g. Masetti et al. \cite{scuba:mas00}; 
Berger et al. \cite{scuba:berger00}).
These variations might be due to refreshing of the shock in the fireball 
(Dai \& Lu \cite{scuba:dl01}), or they may be due to inhomogeneities in 
the ambient medium that the fireball is expanding into 
(Berger et al. \cite{scuba:berger00}).
The achromatic fluctuations have also been explained as a 
micro-lensing event due to a $0.5 M_{\sun}$ lens located half way to 
the burst (Garnavich et al. \cite{scuba:gls00}; but see also
Panaitescu \cite{scuba:pan01}): the peak
magnification of $\sim 2$ occurred 3.8 days after the burst.

Detections of a Lyman alpha break (Smette et al. \cite{scuba:smette01}) 
and weak optical absorption lines (Castro et al. \cite{scuba:castro00a};
Feng et al. \cite{scuba:fww00}; Jensen et al. \cite{scuba:jfg01})
indicated that the burst was at $z = 2.04$.
A faint, extended, $R = 28$ magnitude object was tentatively suggested 
to be the host galaxy (Fruchter \& Vreeswijk \cite{scuba:fv01}).

At the location of the optical transient, a radio and millimeter source was 
found by the VLA, Ryle, Owens Valley, and IRAM 
(Berger et al. \cite{scuba:berger00}).
Our SCUBA results are listed in Table 1.
Unfortunately, our observations on the first day had to be terminated early,
resulting in a large rms.

The SCUBA results for the second day are somewhat confusing.
While it appeared that a source was present, one of our six runs 
had a rather negative result.
There is some indication that the atmosphere was a little unstable:
there were a couple of deviations in the opacity monitored by the 
Caltech Sub-millimeter Observatory.
However, the JCMT was probably looking in a completely different direction 
at the time.
In Table 1, we have conservatively chosen to show the result that uses all 
of our data, including the negative outlier.
If we drop the negative run and combine with the data from the first day,
we instead get a ``best'' result of $3.3 \pm 1.2$ mJy around the time of 
the peak magnification in the lensing scenario.
This is consistent with the other multi-wavelength results that showed that 
the broad-band spectrum was peaking in the sub-millimeter at this time
(Berger et al. \cite{scuba:berger00}).

Our long SCUBA observations on the third day gave a lower rms, but did
not result in a detection of the source at the level suggested from the 
previous day.
This is fully consistent with the results at other wavelengths, that 
showed that the achromatic fluctuation had subsided.
However, given the weak significance of our results, we caution against
over-interpreting the possible transient magnification of the 850 $\mu$m flux.

Combining the data from all three days gives a flux density of
$1.9 \pm 0.7$ mJy.
The rms is the lowest value we have obtained for any burst to date.
In this case, the flux is consistent with coming entirely from the afterglow.
Thus there is no evidence for a quiescent dusty host galaxy for GRB 000301C.
Using the 850 $\mu$m rms value implies a SFR ($M \ge 5 M_{\sun}$)
$< 175 ~M_{\sun} {\rm yr}^{-1}$, assuming $\alpha = 3.4$.
Thus the optical faintness of the host galaxy is unlikely to be explained 
as due to dust absorption in the host; this is consistent with the low 
extinction derived from the optical properties of the afterglow.

\subsection{GRB 000630}

GRB 000630 was detected 20000630.02 by {\it Ulysses}, {\it Wind}, 
{\it NEAR}, and {\it BeppoSAX} (Hurley et al. \cite{scuba:hur00b}).
An unknown radio source was found inside the GRB error box
(Berger \& Frail \cite{scuba:bf00}).
However, a faint fading optical source was also found in the error box at
a different location (Jensen et al. \cite{scuba:jfp00};
Fynbo et al. \cite{scuba:fynbo01a}).
A faint quiescent object was found at this optical location
(Kaplan et al. \cite{scuba:kek01}).

Only one SCUBA observation was attempted on this burst, on July 2,
with the radio source as the target.
The weather was very poor, and no source was detected with an rms of 
2 mJy at 850 $\mu$m.
Since the optical source was the more likely counterpart, we have not
included this result in Table 1.

\subsection{GRB 000911}

The long GRB 000911 was detected 20000911.30 by {\it Ulysses}, 
{\it NEAR}, and {\it Wind} (Hurley et al. \cite{scuba:hur00c}).
A new radio and optical source was found inside the GRB error box
(Berger et al. \cite{scuba:bpf00}).

The optical power law decay was $\delta \sim 1.5$ 
(Price et al. \cite{scuba:pggd00}; Lazzati et al. \cite{scuba:laz00}).
In the simple adiabatic piston model, a fireball produced by a 
one time impulsive injection of energy in which only the forward blast
wave efficiently accelerates particles predicts a power law spectrum 
$S_\nu \propto \nu^{-\beta}$ with
energy spectral index $\beta = 2 \delta / 3$
(Wijers et al. \cite{scuba:wij97}).
For GRB 000911, this would imply $\beta \sim 1.0$.
While the measured broad-band optical spectrum had a steeper $\beta \sim 1.5$
(Lazzati et al. \cite{scuba:laz00}; Covino et al. \cite{scuba:cov00}),
correcting for Galactic extinction brings the spectrum closer to that of
the simple fireball model.

The SCUBA results from our first two observations on the radio-optical 
counterpart are listed in Table 1.
These were originally reported in Smith \& Tilanus (\cite{scuba:st00}).
We did not detect the source at 850 $\mu$m.
A third SCUBA observation that was listed in Smith \& Tilanus
(\cite{scuba:st00}) was made on the incorrect coordinates given in 
Price et al. (\cite{scuba:pgg00}: see Price \cite{scuba:price00}), 
and is not included here.

It is currently difficult to meaningfully compare our SCUBA limits with the 
results of the afterglow emission at other wavelengths.
Simply scaling from the radio data using our fit for GRB 980329 would have 
led to a higher 850 $\mu$m flux than we observed.
However, the GRB 000911 radio data still needs to be 
corrected for possible interstellar scintillation effects.
Similarly, extrapolating the optical spectrum would also have predicted
a higher 850 $\mu$m flux than we observed, although the final 
extinction corrections need to be incorporated.
It does appear, however, that the SCUBA results will ultimately imply there
are at least two breaks in the spectrum between $10^{11}$ and $10^{13}$ Hz, 
as is seen in other afterglows.

Combining all the SCUBA data gives an rms of 0.9 mJy.
Thus there is no evidence for a quiescent dusty host for GRB 000911.
There has also not yet been any indication of a host galaxy in the 
optical observations.
Since no redshift has been reported so far, we are unable to put an
accurate limit on the SFR, although it is likely to be
$< 200 ~M_{\sun} {\rm yr}^{-1}$.
As for GRB 000301C, the optical faintness of the host galaxy is 
unlikely to be explained as due to dust absorption in the host.

\subsection{GRB 000926}

GRB 000926 was detected 20000926.99 by {\it Ulysses}, {\it Wind} 
and {\it NEAR} (Hurley et al. \cite{scuba:hur00d}).
{\it BeppoSAX} and {\it CXO} observations located a fading X-ray 
counterpart in the burst error box that had a very steep power law decay
(Piro \cite{scuba:piro00}; Piro \& Antonelli \cite{scuba:pa00};
Garmire et al. \cite{scuba:gar00}).

Consistent with the location of this X-ray source, a bright new optical and
radio counterpart was found (Gorosabel et al. \cite{scuba:gccg00}; 
Dall et al. \cite{scuba:dall00}; Frail \& Berger \cite{scuba:fb00};
Fynbo et al. \cite{scuba:fynbo01b}).
The optical decay showed a sharp achromatic steepening with time 
(e.g. Fynbo et al. \cite{scuba:fynbo00b}, \cite{scuba:fynbo01b}; 
Rol et al. \cite{scuba:rol00}; Price et al. \cite{scuba:phg01}).
An $R \sim 25$ magnitude compact knot of emission was found at 
this location, which may be the host galaxy
(Harrison et al. \cite{scuba:hgb01}).
A redshift of $z = 2.04$ was determined using the optical absorption 
lines in the afterglow spectrum (Fynbo et al. \cite{scuba:fynbo00c}; 
Castro et al. \cite{scuba:castro00b}).

Unfortunately, GRB 000926 was in the afternoon sky during a period of
poor weather at Mauna Kea.
Only one long SCUBA observation was performed, in very marginal weather.
The result is listed in Table 1.
While the result was tantalizing, it was not possible to obtain a
confirmation.
Extrapolating from the $\sim 0.5$ mJy flux density at 8.46 GHz 
(Frail \& Berger \cite{scuba:fb00}), the 850 $\mu$m flux density
would be expected to be $\sim 1.7$ mJy using a $S_\nu \propto \nu^{1/3}$ 
spectrum, or $\sim 14$ mJy for $S_\nu \propto \nu^{0.9}$ (as in GRB 980329;
see \S1.1).
Thus the SCUBA flux is consistent with coming from the afterglow.

The SFR was estimated to be $14 ~M_{\sun} {\rm yr}^{-1}$ without 
including extinction in the host galaxy, and $24 ~M_{\sun} {\rm yr}^{-1}$
with extinction (Fynbo et al. \cite{scuba:fynbo01b}).
Using the extinction corrected value, the expected quiescent flux 
density at 850 $\mu$m would then be $\lesssim 0.1~{\rm mJy}$, 
assuming $\alpha = 3.4$.
Thus we would not expect to significantly detect a quiescent
sub-millimeter source using SCUBA.

\section{Discussion}

With the possible exception of GRB 980329, for all of the bursts that
were observed by SCUBA between 1997 and 2000, any sub-millimeter emission 
is consistent with coming from the afterglow: we did not conclusively 
detect quiescent sub-millimeter counterparts for any of these sources.

Unfortunately, the redshift remains uncertain for GRB 980329.
If we assume that there is a 1 mJy 850 $\mu$m quiescent source 
present, then we obtain a SFR ($M \ge 5 M_{\sun}$) of 
$240~ M_{\sun} {\rm yr}^{-1}$
if $z = 1$ (using $\alpha = 3.4$), or
$160~ M_{\sun} {\rm yr}^{-1}$
if $z = 5$ (Fruchter \cite{scuba:fruchter99}).
However, since the chance of detecting a random $\ge 1$ mJy 
source in any pointing is $> 10$\%, we cannot rule out that the 
1 mJy quiescent emission that may have been seen in GRB 980329 
is from an unrelated source in the JCMT beam.

Table 2 summarizes the SFR limits for the other bursts that have had 
secure counterpart determinations and that had rms values less than 
2 mJy measured by SCUBA.
For those bursts with reasonably reliable redshift determinations, 
the SFR limit has been estimated.
Our sample spans a wide range of redshifts.
No redshifts have so far been reported for GRB 980519 or 
GRB 000911; however, since the SFR calculations are fairly insensitive 
to $z$, these are likely to have similar SFR limits to the other 
bursts in Table 2.

\begin{table}[t]
\caption[]{Inferred SFR limits ($M \ge 5 M_{\sun}$)
using SCUBA 850 $\mu$m rms values, and $\alpha = 3.4$.}
\label{table2}
\begin{flushleft}
\[
\begin{tabular}{llll}
\hline
\noalign{\smallskip}
Burst   & $z$  & 850 $\mu$m rms & SFR ($M \ge 5 M_{\sun}$)\\
        &      & (mJy)          & ($M_{\sun} {\rm yr}^{-1}$) \\
\noalign{\smallskip}
\hline
\noalign{\smallskip}
971214  & 3.418 & 1.0 & $< 200$ \\
980519  &       & 1.8 &         \\
980703  & 0.966 & 1.6 & $< 380$ \\
990123  & 1.6   & 0.7 & $< 180$ \\
991208  & 0.707 & 1.8 & $< 370$ \\
991216  & 1.0   & 1.2 & $< 290$ \\
000301C & 2.04  & 0.7 & $< 175$ \\
000911  &       & 0.9 &         \\
\noalign{\smallskip}
\hline
\end{tabular}
\]
\end{flushleft}
\end{table}

Radio observations of GRB 980703 that were made years 
after the burst indicated the presence of a persistent radio source
(Berger et al. \cite{scuba:berger01}).
If this radio emission is related to star formation, the inferred
SFR ($M \ge 5 M_{\sun}$) was estimated to be
$\sim 90~ M_{\sun} {\rm yr}^{-1}$.
Table 2 shows that our 850 $\mu$m SCUBA limit is consistent
with this result.
Our 450 $\mu$m rms of 20 mJy for GRB 980703 places a slightly higher limit
of SFR ($M \ge 5 M_{\sun}$) $< 550~ M_{\sun} {\rm yr}^{-1}$.

The fact that we are not consistently detecting bright sub-millimeter 
sources with SCUBA naturally leads to questions regarding the connection 
between GRBs and dust-enshrouded star formation.
If GRBs are due to the explosions of high-mass stars, this may indicate 
that the relatively small population of extremely luminous dusty galaxies 
does not dominate the total star formation in the universe at early epochs.
Instead, the GRBs may be predominantly tracing slightly lower luminosity 
galaxies with SFR ($M \ge 5 M_{\sun}$)
$\sim 10 - 100 ~M_{\sun} {\rm yr}^{-1}$.

We caution that our sample of bursts is still rather small.
We also note that there may be an important selection effect in the 
bursts studied so far.
Since the current GRB error boxes are large, we have to wait for the 
determination of a counterpart at other wavelengths before proceeding 
with our observations.
If a bright optical afterglow is found, this suggests that the absorption 
local to the source is not too high, and so it may not be too surprising if 
we do not see a quiescent sub-millimeter source.
Better candidates are those in which a radio transient is found, but
little or no optical emission is seen, such as for GRB 980329.
However, optical transients were found for the 6 new bursts studied
in this paper.

This situation should improve significantly with the bursts localized by
{\it HETE-2} and {\it Swift}.
{\it HETE-2} should provide a few bursts each year with positions accurate 
to $\pm 10\arcsec$.
For these, we will be able to immediately use SCUBA in the photometry mode 
on these locations, even if no counterpart is found at other wavelengths.
Thus we will be able to look for prompt, afterglow, and quiescent sources
that may be hard to detect otherwise if the redshift is large.
Sub-millimeter observations performed within a day of the burst can
potentially discriminate between different afterglow models
(Panaitescu \& Kumar \cite{scuba:pk00}; Livio \& Waxman \cite{scuba:lw00}).
It is also exciting that {\it HETE-2} will be able to localize bursts that
last less than 1 second: these may have different progenitors and 
counterpart behaviors from the objects studied to date 
(Kouveliotou et al. \cite{scuba:kouv93}).

Observations of new bursts are continuing to produce surprises, and 
there is much left to learn about GRB afterglows and host galaxies. 
To obtain a complete picture of their nature will require the careful 
study of many bursts to expand our sample.
Sub-millimeter observations with a $\sim$ mJy sensitivity are a key 
component to the multi-wavelength coverage.
To this end, our program of Target of Opportunity observations using 
SCUBA is ongoing.

\begin{acknowledgements}

The James Clerk Maxwell Telescope is operated by The Joint Astronomy 
Centre on behalf of the Particle Physics and Astronomy Research Council 
of the United Kingdom, the Netherlands Organisation for Scientific Research, 
and the National Research Council of Canada.

We thank the JCMT Director Ian Robson for authorizing the ToO observations,
and Graeme Watt, Gerald Moriarty-Schieven, Michiel Reuland, Elese Archibald,
Tim Pickering, Nick Jessop, Iain Coulson, 
and the dedicated efforts of the JCMT telescope operators
for their valuable assistance with the observations and reductions.
We thank Andrew Fruchter and the referee for some helpful suggestions.

We are grateful to Scott Barthelmy and Paul Butterworth of 
The GRB Coordinates Network (GCN), and to the other ground-based 
observers for the rapid dissemination of their burst results.

This work was supported by NASA grants NAG 5-3824 and 5-9904
at Rice University.

\end{acknowledgements}

\end{document}